\documentclass[aps,prd,twocolumn,nofootinbib]{revtex4}
\usepackage{amssymb}
\usepackage{amsmath,color}
\usepackage{epsfig}
\usepackage{graphicx,epsf,epsfig}
\usepackage{bbm}
\usepackage{subfigure}
\usepackage{multirow}
\usepackage{setspace}
\usepackage{verbatim}
\usepackage[utf8x]{inputenc}
\usepackage{epstopdf}
\usepackage[linktocpage]{hyperref}
\usepackage{multirow}
\newcommand{\be}{\begin{eqnarray}}
	\newcommand{\ee}{\end{eqnarray}}
\newcommand{\bq}{\begin{equation}}
	\newcommand{\eq}{\end{equation}}
\newcommand{\bp}{\begin{split}}
	\newcommand{\ep}{\end{split}}

\begin{document}
	
	\title{NANOGrav Signal from the End of Inflation and the LIGO Mass \\ and Heavier Primordial Black Holes}
	
	\author{Amjad Ashoorioon}
	\email{amjad@ipm.ir}
	
	\author{Kazem Rezazadeh}
	\email{kazem.rezazadeh@ipm.ir}
	
	\author{Abasalt Rostami}
	\email{aba-rostami@ipm.ir}

	\affiliation{School of Physics, Institute for Research in Fundamental Sciences (IPM), 	P.O. Box 19395-5531, Tehran, Iran}

	\begin{abstract}
		
		Releasing the 12.5-year pulsar timing array data, the North American Nanohertz Observatory for Gravitational Waves (NANOGrav) has recently reported the evidence for a stochastic common-spectrum which would herald the detection of a stochastic gravitational wave background (SGWB) for the first time. We investigate if the signal could be generated from the end of a $\sim 10$ MeV but still phenomenologically viable double-field inflation when the field configuration settles to its true vacuum. During the double-field inflation at such scales, bubbles of true vacuum that can collapse to LIGO mass and heavier primordial black holes form.  We show that only when this process happens with a first-order phase transition, the produced gravitational wave spectrum can match with the NANOGrav acclaimed SGWB signal.   We show that the produced gravitational wave spectrum matches the NANOGrav SGWB signal only when this process happens through a first-order phase transition. Using LATTICEEASY, we also examine the previous observation in the literature that by lowering the scale of preheating,  despite the shift of the peak frequency of the gravitational wave profile to smaller values, the amplitude of the SGWB could be kept almost constant. We notice that this observation breaks down at the preheating scale, $M\lesssim 10^{-14}~m_{{}_{\rm Pl}}$.
		
	\end{abstract}
	
	\maketitle
	
	\section{Introduction}
	\label{sec:int}
	
	Following a series of striking discoveries of gravitational wave (GW) signals from mergers of binary black holes, by LIGO and VIRGO groups  \cite{LIGOScientific:2016aoc}, binary neutron stars \cite{LIGOScientific:2017ync}, and recently black hole-neutron star mergers \cite{LIGOScientific:2021qlt}, the North American Nanohertz Observatory for Gravitational Waves (NANOGrav) has reported its $12.5$-year data set by scrutinizing the cross-power spectrum of pulsar timing residuals \cite{NANOGrav:2020bcs}. The group has announced signs of a stochastic common-spectrum process, parameterized in a power-law form. However, to judge that the detected signal is an astrophysical or cosmological stochastic GW background, one needs to discriminate monopolar, dipolar, and quadrupolar correlation signatures.
	
	Although the nature of the NANOGrav is yet to be confirmed,  due to a host of consequences that such a signal might have for the early universe cosmology, it is worth considering the signal seriously and proposing possible mechanisms that could create such a signal. It has also been argued that the NANOGrav signal could be the second-order GWs associated with the formation of solar-mass primordial black holes (PBHs) \cite{Kohri:2020qqd}. To match with the slope of the secondary SGWB, there must be a dust-like post-inflationary stage before radiation dominated era which suggests a considerable existence of planet-mass primordial black holes \cite{Domenech:2020ers}.  Alternatively, \cite{Vaskonen:2020lbd} claims that the NANOGrav signal could be attributed to a stochastic gravitational wave signal associated with the formation of supermassive primordial black holes from high-amplitude curvature perturbations. Such high amplitude curvature perturbations could be generated during inflation, see e.g. \cite{Ashoorioon:2019xqc}. In the context of supersymmetric inflation, such a secondary signal was argued to be the signal from a peak in the power spectrum which would have led to the formation of PBHs in the mass range $10^{-12}-10^{-6}\, M_{\odot}$ at the price of lowering the black hole abundance to the level which cannot explain the dark matter energy density \cite{Spanos:2021hpk} (see \cite{Ahmed:2021ucx}) for a non-supersymmetric inflationary model). In \cite{Nakai:2020oit}, the authors achieve a NANOGrav signal, if there was a 1PT  in the dark sector around a few GeV, which is assumed to be completely decoupled from the visible sector except for gravitational interaction. The phase transition was claimed to mitigate the outstanding Hubble tension. It has also been shown that an inflationary tensor power spectrum can lead to such a signal if the spectrum of the inflationary produced gravitational waves is blue, $0.7\lesssim n_{{}_{\rm T}}\lesssim1.3$,  $r\gtrsim 10^{-6}$, and the reheating temperature is roughly smaller than $100-1000$ GeV \cite{Vagnozzi:2020gtf}. Inflationary models, however, tend to always produce a red tensor power spectrum as long as the inflaton energy-momentum tensor satisfies the null energy condition unless the initial condition for tensor perturbations is scale-dependent excited state \cite{Ashoorioon:2014nta}. Also, a blue pure power-law gravitational wave, compatible with the latest bounds on the CMB scales, which extrapolates to NANOGrav frequencies would conflict with the upper limits on the stochastic GW background (SGWB) amplitude from LIGO/Virgo. Ref. \cite{Benetti:2021uea}, hence considered a broken power-law form for the SGWB to overcome this problem. It has also been claimed that the NANOGrav signal can come from the collapse of closed domain walls generated during inflation at the stage of occurrence of sphericity \cite{Sakharov:2021dim}. Finally, \cite{Pandey:2020gjy} argues that NANOGrav GWs could be generated due to the instability caused by the finite difference in the number densities of the different species of the neutrinos in a hot dense neutrino asymmetric plasma. To explain the signal, the magnetic field strength should be at least  $\sim 10^{-12}$ G at the Mpc length scale.
	
	Since rolling and phase transition is quite ubiquitous in the string theory landscape \cite{Susskind:2003kw}, it is also plausible to consider the possibility that a NANOGrav signature might be caused by a 1PT which immediately occurred after a rolling phase, which served as the primordial inflation \cite{Adams:1990ds,Ashoorioon:2015hya,Ashoorioon:2020hln}. In this paper, we work out a concrete double-field inflationary model with enough reheating temperature suitable for nucleosynthesis, in which rolling inflation ends via slow-roll violation before the universe settles from the metastable valley by a first-order phase transition. In fact, in the course of inflation,  the presence of another direction with smaller vacuum energy causes the bubbles of true vacuum with small radii, often smaller than the Hubble radius in the true vacuum $\sim H_t^{-1}$ to form and then stretch to larger lengths due to the inflationary dynamics. Some of these bubbles become even larger than $H_t^{-1}$ and due to the vacuum energy inside the bubble, they keep inflating. This leads to an inflating baby universe in the post-inflation era, which is connected to the parent radiation-dominated FRW universe through a wormhole. Such wormholes pinch-off in a time scale, $t_{{}_{\rm pinch}}\sim H_t^{-1}$, and {\it supercritical} PBHs with mass of $\mathcal{O}(t_{{}_{\rm pinch}}\, M_{{}_{\rm Pl}}^2)$ form in the mouth of two throats \cite{Deng:2017uwc}. On the other hand, the {\it subcritical} bubbles with a radius smaller than $\sim H_t^{-1}$  lose their energy after inflation by interacting with radiation in a very short time scale. They recede from the cosmological expansion and collapse to form black holes \cite{Deng:2017uwc}. The mass function of PBHs in this scenario strongly depends on the time span after the end of inflation, $\Delta t_p$, during which the 1PT completes \cite{Ashoorioon:2020hln}. If $\Delta t_p < t_{{}_{\rm pinch}}$, as the large bubbles collide and percolate, only the subcritical bubbles find the opportunity to collapse and the supercritical PBHs have no significant contribution to the mass fraction of PBHs. We will see that in the scenario that we propose the NANOGrav SGWB signal from the first-order phase transition after the end of inflation,  bubbles of true vacuum form that can collapse to PBHs of a few tens to few ten thousand of the solar mass, depending on the exact value of Hubble parameter during inflation. The abundance of the PBHs in this mass range is proportional to the probability of nucleation from the metastable direction to the true vacuum. Hence by measuring the abundance of such PBHs, we could in principle chart the landscape in our neighborhood \cite{Ashoorioon:2020hln}.
	

	The SGWB spectrum can also be produced during the preheating phase in which the universe exits the inflationary stage by a second-order phase transition (2PT) and starts oscillating at the bottom of the potential. Coupling of the inflaton to other fields, known as preheating fields,  parametrically excites the preheat fields in some instability bands in momentum space, which are equivalent to inhomogeneities in the position space. Such inhomogeneities source the tensor perturbations and can produce a stochastic background of the gravitational wave spectrum. The scattering of these excited modes with the rest finally reheats and thermalizes the universe. The production of gravitational waves from the preheating phase was first investigated by \cite{Khlebnikov:1997di}. The subject was further studied in a variety of inflationary settings \cite{Easther:2006vd, Easther:2006gt, Easther:2007vj, Dufaux:2007pt, Figueroa:2017vfa, Garcia-Bellido:2002fsq, Garcia-Bellido:2007nns, Garcia-Bellido:2007fiu, Dufaux:2008dn, Felder:2000hj, Felder:2001kt, Cui:2021are, Antusch:2016con, Amin:2018xfe, Lozanov:2019ylm, Hiramatsu:2020obh, Kou:2021bij, Ashoorioon:2013oha}. Hence, one may wonder if the exit from the metastable direction and stochastic resonance around the true vacuum in such a low-scale double-field inflationary model, can also yield a stochastic gravitational wave background that resembles the NANOGrav signal. Especially in \cite{Easther:2006vd}, the authors focused on numerical calculations of the GW spectrum from preheating in inflationary models with energy scales much lower than the GUT scale by solving for the equations of the metric perturbations in the Fourier space. For this purpose, they implemented a lattice simulation for an effective potential in the simple quadratic form of $V(\phi)=\mu^{2}\phi^{2}/2$, which they assume to be the approximation of the potential during preheating. They concluded that for all the masses in the range $\mu\sim10^{-18}-10^{-6}m_{{}_{\rm Pl}}$, the amplitude of the induced GWs is the same and of order $\Omega_{{}_{\rm GW}}h^{2}\sim10^{-11}-10^{-10}$, but the relevant frequencies of the gravitational spectrum shift to lower values with the reduction of $\mu$. This motivated us to examine the generation of the stochastic GW background from preheating in our double-field inflationary scenario, where the inflaton settles to the true vacuum by a tachyonic instability, and preheating occurs through stochastic resonance. For this purpose, we utilized the LATTICEEASY code \cite{Felder:2000hq} which is publicly available. Implementing a lattice simulation, we estimate the amplitude of GWs produced from the preheating process in our scenario. In particular, we examine if it is possible to explain the observed NANOGrav signal from a second-order phase transition through the mechanism explained earlier. We find that the produced signal is too weak, $\Omega_{{}_{\rm GW}}h^{2}\lesssim10^{-77}$, which is well below the acclaimed NANOGrav signal, and in a completely different frequency band, $10^2~{\rm Hz} \lesssim f \lesssim10^4$ Hz. We also investigate this question in the context of the model proposed in \cite{Easther:2006vd}, even though the model is not consistent with the PLANCK 2018 constraints on the CMB scales \cite{Planck:2018jri}. We noticed that the observation of \cite{Easther:2006vd} is valid until $\mu\sim 10^{-26} m_{{}_{\rm Pl}}$, which corresponds to preheating energy scale of $M\gtrsim 10^{-14}~m_{_{{\rm Pl}}}$. Below this value, the resulting stochastic gravitational wave background spectrum does not remain a fixed fraction of order $10^{-11}-10^{-10}$ of the critical energy density. With lowering the energy density of the inflaton at the beginning of the preheating from $\mu\sim 10^{-26} m_{{}_{\rm Pl}}$, the amplitude of the produced GWs starts to diminish like $\mu^4$.
	

	The paper is organized as follows: In section \ref{sec:PBH}, we first construct a double-field inflationary model in which inflation ends via slow-roll violation before making a first-order phase transition to the true vacuum. As will be shown in section \ref{sec:results}, the model is not only compatible with the CMB observations but also produces a gravitational wave signal compatible with NANOGrav analysis. In section \ref{section:preheating}, we also examine the possibility of generation of the NANOGrav signal from a second-order phase transition and through the preheating process after inflation in our two-field scenario. In section \ref{PBHs}, it is shown how a correlated signal, as PBHs in the mass range of $30-36000~M_{{}_{\odot}}$  could be generated from the collapse of subcritical bubbles produced during the course of double-field inflation. Finally, in section \ref{conclusions}, we summarize our results and conclude the paper.

	\section{The Double-Field Model}
	\label{sec:PBH}
	
	The natural synthesis of old and new inflation is combined in the context of ``double-field'' inflation in which while the waterfall field $\psi$ is initially trapped in its meta-stable (false) vacuum, the slowly rolling inflaton $\phi$ drives inflation. As inflation progresses, the nucleation rate of false to true vacuum transition gradually grows and eventually becomes significantly large at some critical field value $\phi_{\rm{pt}}$ so that the bubbles of true vacuum can percolate \cite{Adams:1990ds, Ashoorioon:2015hya, Ashoorioon:2020hln}. This is similar to the standard Hybrid inflationary models with this difference that inflation terminates by a 1PT rather than tachyonic instability of waterfall field \cite{Linde:1993cn}. In general, the formal potential of double-field inflation can be written as
	\be\label{pot-tot}
	V(\phi,\psi)=V_0+V_1(\phi)+ V_2(\phi,\psi)\,,
	\ee
	where the detailed dynamics of the inflation is mainly governed by the inflaton potential $V_1$ and the constant vacuum energy $V_0$, while $V_2$ comes into play when the newly developed minimum in $\psi$ direction drops effectively below the first one, which is occupied during inflation. Here, we assume a slightly different extended hybrid inflationary potential \cite{Copeland:1994vg}, in which the dynamics of the slowly rolling inflaton is given by an inflection point inflation potential
	\begin{eqnarray}
		&&V(\phi,\psi)=V_0+\underbrace{\frac{1}{2} m^2{\phi}^2 - \frac{ A \lambda_{{}_3}}{3} {\phi}^3 + \lambda_{{}_3}^2 {\phi}^4}_{V_1(\phi)}+  \nonumber\\
		&& \underbrace{\frac{1}{4}\lambda \psi^4-\frac{1}{3}\gamma M \psi^3 +\frac{1}{2} \lambda^{\prime} \phi^2\psi^2+\frac{1}{2} \alpha M^2 \psi^2}_{V_2(\phi,\psi)}\,.
		\label{Potential}
	\end{eqnarray}
	From the above explicit form of potential, it is clear that for large values of $\phi$ the global minimum of $V_2(\phi,\psi)$ is located at $\psi_{{}_{\rm{min1}}}=0$. If the field value becomes less than
	\be
	\phi_{{}_{\rm I}}^2=M^2 \frac{\gamma^2-4\alpha \lambda}{4\lambda^{\prime}\lambda}\,,
	\ee
	then the potential develops a second minimum in $\psi$ direction at
	\begin{align*}
		\psi_{{}_{\rm{min2}}} = \frac{M\gamma + \sqrt{M^2\gamma^2 - 4 M^2\alpha\lambda - 4 \lambda\lambda' \phi^2}}{2\lambda}\,.
	\end{align*}
	The transition between two minima occurs when the newly-developed one drops below the first occupied one. This happens for the above potential for $\phi < \phi_{{}_{\rm cr}}$, where
	\begin{equation}
		\phi_{{}_{\rm cr}}\equiv \frac{2\gamma^2M^2}{9\lambda\lambda'} - \frac{\alpha M^2}{\lambda'}\,,
	\end{equation}
	is the critical value of the inflaton field for which degeneracy of minima of $V_2$ takes place. In principle, by judicious choice of the inflaton potential $V_1(\phi)$ in Eq.(\ref{Potential}), one can simply adjust the predictions of the model at cosmological scales to agree with the CMB observables. As will be discussed later on, to produce the NANOGrav gravitational wave signal, we need a very low-scale inflationary period with a nearly constant Hubble parameter of order $\mathcal{O}(10^{-41})M_{\rm{Pl}}$. It is convenient to exploit the inflection point inflationary potential of the following form to satisfy such a delicate requirement,
	\be\label{Inflec-Pot}
	V_1(\phi)&=& \frac{1}{2} m^2{\phi}^2 - \frac{A \lambda_3}{3} {\phi}^3 + \lambda_3^2 \phi^4\,.
	\ee
	The general form of this potential could be realized in the context of Matrix inflation \cite{Ashoorioon:2009wa}. The most important point about this choice of inflationary potential is that one can always lower the inflation energy scale to the desired level, whilst the requirement of the amplitude of the density perturbations and the spectral tilt at the CMB scales are kept in agreement with the observations. In the first version of (\ref{Inflec-Pot}), the parameters were chosen such that inflation just comes about in the neighborhood of the inflection point, say $\phi_0$. For the most vanilla inflection model, where the first and second derivatives of Eq. (\ref{Inflec-Pot}) vanish at the inflection point, the minimum number of e-folds required for an inflationary model at few MeV scale is $N_e\approx 20$ for which the scalar spectral index is  $n_{{}_S} \simeq 1-4/N_e=0.80$, which is well outside of the Planck 2018 95\% C.L. region \cite{Planck:2018jri} \footnote{The required number of e-foldings to solve the problems of the Standard Big Bang cosmology for an inflationary model at the energy scale $M$ and reheating temperature $T_{\ast}$ is given by $N_e=53+\frac{2}{3}\ln(M/10^{14}{\rm GeV})+\frac{1}{3}\ln(T_{\ast}/10^{10}{\rm GeV})$}.  To make the model more flexible, one may perturb the parameter $A$ using a new small dimensionless parameter $\nu$ as
	\be\label{A-pert}
	A = 4m\left(1-\frac{1}{4}\nu^2\right)^{\frac{1}{2}}\,,
	\ee
	which in turn shifts the inflection point \footnote{It is easy to see that $V'_1(\phi_i)\neq 0$ although $V''_1(\phi_i)=0$}  to
	\be
	\phi_i = \phi_0 \Big(1 - \frac{1}{4}\nu^2 +\mathcal{O}(\nu^4)\Big)\,.
	\ee
	If inflation occurs in some vicinity of this new inflection point, the modified version of (\ref{Inflec-Pot}) can be written down as
	\begin{eqnarray}\label{inflection-point}
		V_1(\phi) &\approx& \frac{1}{12} m^2\phi_0^2\Big(1+\nu^2\Big) + \frac{\nu^2}{4}m^2\phi_0\Big(\phi-\phi_i\Big)\\ \nonumber
		&+&\frac{2m^2-m^2\nu^2}{6\phi_0}\Big(\phi-\phi_i\Big)^3\,,
	\end{eqnarray}
	where $m, \phi_0$, and $\nu$ may be determined if we impose not only the compatibility of the model with the CMB observations at cosmological scales but also the negligibility of the vacuum energy $V_0$ in comparison with $V_1$ in the course of inflation (i.e., $|V_0|\ll |V_1|$). Moreover, demanding a zero vacuum energy after the phase transition relates the constant $V_0$ to the parameter space of $V_2$ at the global minimum  of (\ref{Potential})
	\be
	V_2(\psi_{{}_{\rm{min2}}},\phi=0)=-V_0\,.
	\label{zero-constant}
	\ee
	Using (\ref{inflection-point}), one can simply find the following relations between the parameters space and the relevant observational quantities at the CMB scales, namely the power spectrum $\mathcal{P}_{{}_S}$ and spectral index $n_{{}_S}$ in term of the dimensionless parameter $\beta=6\nu/\phi_0^2$
	\begin{eqnarray}\label{obser-para}
		n_{{}_S} &=& 1-4\beta\, \cot\beta N_e\,,\\ \nonumber
		\mathcal{P}_{{}_S}^{1/2}&\approx& 0.625\, \frac{m}{\phi_0}\Big(\frac{\sin^2\beta N_e}{\beta^2}\Big)\,.
	\end{eqnarray}
	Since inflation happens near the inflection point, the Hubble parameter would be approximately specified with $H^2 \approx m^2\phi_0^2/36$. This, along with the relations in (\ref{obser-para}), uniquely determines the parameters space of (\ref{inflection-point}), for any chosen value of $H$. For instance, in the case of first-order phase transition at approximately $1-20$ MeV that can explain the NANOGrav signal, the Hubble parameter would be about $H\approx 10^{-40}M_{{}_{\rm{Pl}}}$ for which the required number of e-folds becomes $N_{\rm{e}}\approx 20$. For example, we take the following values of parameters,
	\begin{eqnarray}\label{inf-data}
		m&=&5.59569\times 10^{-29}M_{{}_{\rm{Pl}}}\,, \nonumber\\   \phi_0&=&8.5431144\times 10^{-12}M_{{}_{\rm{Pl}}}\, ,  \\  \nu&=&8.823552\times 10^{-25}\,,\nonumber
	\end{eqnarray}
	to realize the observed power spectrum $\mathcal{P}_{{}_S}\approx 2\times 10^{-9}$ and spectral tilt $n_{{}_S} \approx 0.965$. For these values of parameters, if inflation ends with slow-roll violation $\epsilon \approx 1$, then the points $\phi_{{}_{\rm CMB}}$ and $\phi_{{}_{\rm e}}$ at which the observable scales leave the horizon and inflation terminates respectively, are given by
	\begin{eqnarray}
		&&\phi_{{}_{\rm e}} = 8.543105\times 10^{-12}M_{{}_{\rm{Pl}}},\\ \nonumber
		&&\phi_{{}_{\rm CMB}} \simeq \phi_0\,.
	\end{eqnarray}
	In the next sections, we investigate different scenarios for settling from the false valley to the true vacuum.
	
	\section{First-Order Phase Transition after the End Of Inflation and NANOGrav Gravitational Waves}
	\label{sec:results}
	
	When the end of inflation comes about by the slow-roll violation, sometime before the 1PT,  the real reheating process does not take place until the 1PT  completes through bubble collision. One can assume that the universe is cold after inflation at the onset of 1PT and apply the formalism and equations of generation of a stochastic background of gravitational waves at zero temperature from bubble collision during a first-order phase transition. Ref. \cite{Kosowsky:1992vn} for the first time computed the profile of gravitational waves from colliding vacuum bubbles using a combination of envelope approximation and simulations of hundreds of bubbles. Later Ref. \cite{Huber:2008hg} used a larger number of bubbles with the envelope approximation. The spectrum takes an asymmetric  dome in the vicinity of the peak frequency $f_{{}_{\rm m}}$, given by
	\be\label{frequency-peak}
	f_{{}_{\rm m}}=3\times 10^{-8} {\left(\frac{g_{{}_{\ast}}}{100}\right)}^{1/6}  \left(\frac{T_{{}_\ast}}{1  \rm GeV}\right) \left(\frac{\beta}{H_{{}_{\rm pt}}}\right)\,.
	\ee
	
	\begin{figure}
		\begin{center}
			\scalebox{0.65}[0.65]{\includegraphics{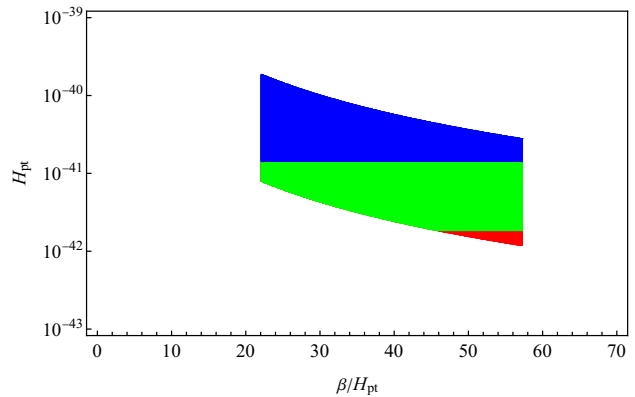}}
			\caption{The parameter space of $H_{{}_{\rm pt}}$ versus $\beta/H_{{}_{\rm pt}}$ in our double-field scenario which satisfies the conditions $2.5\times10^{-9}~{\rm Hz}<f_{\rm m}<1.2\times10^{-8}~{\rm Hz}$, $3\times10^{-10}<\Omega_{{}_{\rm GW}}h^{2}<2\times10^{-9}$ and $\beta/H_{{}_{\mathrm{pt}}}>1$ together with the different lower bounds on the reheating temperature. The union of blue, green, and red regions is compatible with the lower bound $T>1\,\mathrm{MeV}$, and the union of blue and green regions is compatible with $T>1.8\,\mathrm{MeV}$, whereas the blue region is only compatible with $T>5\,\mathrm{MeV}$.}
			\label{figure:H-betaH}
		\end{center}
	\end{figure}
	
	
	\begin{figure}
		\begin{center}
			\scalebox{0.7}[0.7]{\includegraphics{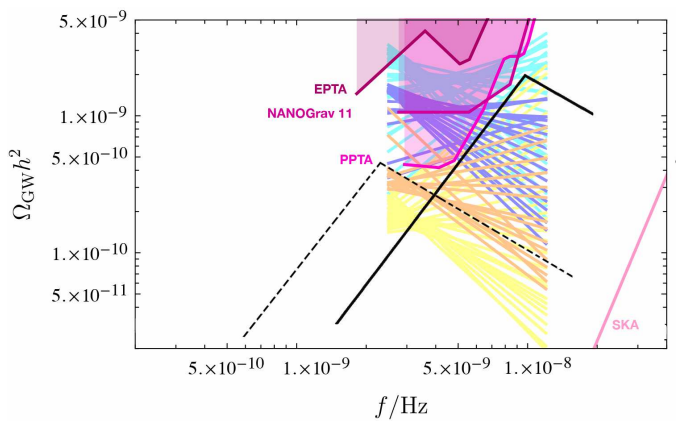}}
			\caption{GW spectra produced from the 1PT leading to PBHs of mass $330- 2700 M_{\odot}$ (thick black curve). Regarding an approximated visual guide of the NANOGrav signal range, we have drawn power-law lines in the interval $2.5\times 10^{-9} \leq f \leq 1.2\times 10^{-8}$Hz. The blue (cyan) and orange (yellow) lines are on the upper and lower $1\sigma$ ($2\sigma$) contours of the NANOGrav signal, respectively. The dashed line, on the other hand, shows the case in which the peak frequency and its corresponding GW amplitude from 1PT are $f_m \approx 2.4\times 10^{-9}$ and $\Omega_{{}_{\rm GW}} \approx 5\times 10^{-10}$ and hence out of the NANOGrav region. However, the falling $f^{-1}$ tail passes through the NANOGrav legitimate range. For such 1PT, the Hubble parameter takes the value $H_{{}_{\rm pt}} \approx 9.2 \times 10^{-42} M_{{}_{\rm Pl}}$ and therefore the reheating temperature remains above the nucleosynthesis temperature, $\sim 1.8 $ MeV.}
			\label{omplot}
		\end{center}
	\end{figure}
	
	The gravitational wave amplitude proportionally increases  with $f^{2.8}$ and decreases with $f^{-1}$ for regions $f<f_{{}_{\rm m}}$ and $f>f_{{}_{\rm m}}$ respectively. In (\ref{frequency-peak}), $g_{{}_*}\approx 106$ is the relativistic degrees of freedom, $H_{{}_{\rm{pt}}}$ is the Hubble parameter at the time of 1PT and, the instantaneous reheating temperature $T_{{}_*}$ and $\beta$ is given by
	\be\label{temperature}
	T_{{}_*} &=& \Big(\frac{90 H_{{}_{\rm pt}}^2 M_{{}_{\rm Pl}}^2}{\pi^2  g_{{}_*}}\Big)^{1/4}\,, \\ \nonumber
	\beta &=& \frac{dS_{{}_{\rm E}}}{dt}\,,
	\ee
	in which $\beta^{-1}$ is roughly the timescale that takes for the 1PT to complete. The amplitude of the resulting gravitational wave today can also be evaluated at the value of its peak frequency $f_{{}_{\rm m}}$ as
	\be\label{GW-amplitude}
	\Omega_{{}_{\rm GW}} h^2 (f_{{}_{\rm m}})=10^{-6} \left(\frac{g_{{}_{\ast}}}{100}\right)^{-1/3} {\left(\frac{H_{{}_{\rm pt}}}{\beta}\right)}^2\, .
	\ee
	Now let us assume that the NANOGrav SGWB reported by Ref. \cite{NANOGrav:2020bcs}  within the frequency range $2.5\times 10^{-9}-1.2\times 10^{-8}$ Hz and amplitude $3\times 10^{-10}-2\times 10^{-9}$, is the gravitational wave signal generated after the end of inflation (which happened through slow-roll violation), in the above setup. Obviously, since we are demanding a 1PT right after an inflection point inflation, the Hubble parameter would not change considerably and the quantities $\beta$ and $H_{{}_{\rm pt}}\approx H_{{}_{\rm f}}$ can be determined uniquely by using Eqs. (\ref{frequency-peak})-(\ref{GW-amplitude}) if one is given the peak frequency and corresponding SGWB amplitude.
	Exploiting Eqs. (\ref{frequency-peak}, \ref{temperature}, \ref{GW-amplitude}), the current allowed interval for NANOGrav signals, and also imposing the constraint after the first-order phase transition needs to be above the nucleosynthesis temperature, $\sim 1$ MeV, the Hubble parameter is confined to be in the interval $6.00 \times 10^{-42}\lesssim H_{\rm pt} \lesssim 9.22\times 10^{-40} M_{\rm Pl}$. In Fig. \ref{figure:H-betaH}, we applied the constraints $2.5\times10^{-9}\,{\rm Hz}<f_{{}_{\rm m}}<1.2\times10^{-8}\,{\rm Hz}$, and $3\times10^{-10}<\Omega_{{}_{\rm GW}}h^{2}<2\times10^{-9}$ together with different lower bounds on the reheating temperature, and illustrated the allowed regions for the parameter space of our model in the plane of $H_{{}_{\mathrm{pt}}}$ with respect to $\beta/H_{{}_{\mathrm{pt}}}$. In this graph, the union of blue, green, and red regions specifies the parameter space which is consistent with the lower bound on the reheating temperature as $T>1\,\mathrm{MeV}$. With the assumption of thermalization of the long-lived massive particles, \cite{Hasegawa:2019jsa} computes this lower bound on reheating temperature increases to $1.8\,\mathrm{MeV}$. The allowed parameter space is then restricted to the regions which are specified by blue and green colors. With the assumption of hadronic decay of long-lived massive particles in the mass range 10 GeV to 100 TeV, \cite{Hasegawa:2019jsa} obtain that the minimum on reheating temperature increases to $5\,\mathrm{MeV}$, and with this constraint, the allowed parameter space in our model is restricted to only the blue region in the figure. The maximum reheating temperature in our models could be as high as 17 MeV, which easily satisfies such lower limits on reheating temperature. Another notable thing is that the confined region of parameter space, naturally satisfied $\beta/H_{{}_{\mathrm{pt}}}\gg 1$, which is required for the phase transition to complete in much less than the Hubble time, a necessity imposed to be able to use the results of GW simulations from bubble collision in flat spacetime.
	
	In our setup, for the NANOGrav frequency and amplitude $f_{{}_{\rm m}}\approx 8.81 \times 10^{-9}$Hz, $\Omega_{{}_{\rm GW}} h^2 (f_{{}_{\rm m}})\approx 1.9\times 10^{-9}$, we find $\beta/H_{{}_{\rm pt}}\approx 22.3$ and $H_{{}_{\rm pt}}\approx 9.77\times 10^{-41} M_{{}_{\rm{Pl}}}$ (see Fig.\ref{omplot}). For the 1PT to happen in consistency with these results, we use the parameter set given in (\ref{inf-data}) and also
	\be\label{data-A}
	\lambda&=&1.6708\times 10^{-15}\,,\, \, \gamma=5.0249\times 10^{-21}\,,\\ \nonumber
	\lambda'&=&\alpha=1.0\times 10^{-41}\,,\, \, M=1.14\times 10^{-11} M_{{}_{\rm{Pl}}}\,,\\\nonumber
	V_0&=&1.92\times 10^{-82} M_{{}_{\rm{Pl}}}^4\,,
	\ee
	for the parameters of the potential (\ref{pot-tot}). Due to the reasons that will be mentioned later, we have tuned the parameters of the setup such that the tunneling probability during inflation remains fairly small, which in this case turns out to be around $p\approx 10^{-20}$. After this epoch, it grows exponentially and finally meets its critical value $p_c$ at $\phi_{{}_{\rm pt}} \approx 7.351\times 10^{-24} M_{{}_{\rm Pl}}$. From Eq. (\ref{temperature}), one easily notices that the reheating temperature in the above 1PT takes the value $T_{{}_*} \approx 13$ Mev, which is well above the what is needed for nucleosynthesis.
	
	It is also possible that the NANOGrav probe has only detected the falling tail of the SGWB spectrum generated from the first-order phase transition after inflation.  We have presented a benchmark for this scenario by setting the Hubble scale at 1PT as $H_{{}_{\rm pt}}\approx 9.2 \times 10^{-42}M_{{}_{\rm{Pl}}}$. From Fig. \ref{omplot} it can be seen that although the peak frequency is out of the desired range, the $f^{-1}$ falling tail still passes through the NANOGrav region. The reheating temperature in this case is $1.8$ MeV.
	
	\section{Second-order phase transition and preheating}
	\label{section:preheating}
	
	One may wonder if a similar SGWB could be generated from the parametric resonance at the end of inflation \cite{Kofman:1994rk, Kofman:1997yn}. The spectrum of SGWB from preheating, more or less has an asymmetric $\Lambda$ shape similar to the spectrum of gravitational waves from a first-order phase transition. Ref. \cite{Easther:2006vd} has also claimed that even at small energy scales, the SGWB from preheating can be of order $\Omega_{{}_{\rm GW}}h^2\sim 10^{-11}-10^{-10}$, which is close enough to the amplitude of NANOGrav signal. In this section, we study the production of GWs from the preheating process in our double-field inflationary model, if the process of settling to the true vacuum happens through a second-order phase transition and later preheating from the stochastic resonance occurs from the coupling of the inflaton to preheat fields while the inflaton oscillates around its true minimum $(\phi=0,\psi=\psi_{{}_{\rm min2}})$. To realize this scenario within the setup of potential \eqref{Potential}, we have to assume that $\alpha<0$. The field value at which the potential becomes tachyonic is
	\begin{equation}
		\label{phi_inst}
		\phi_{{}_{\rm inst.}}=M\sqrt{-\frac{\alpha}{\lambda'}}\,,
	\end{equation}
	which we take to be equal or smaller than $\phi_{{}_{\rm e}}$, where inflation ends. We also assume that when $\psi$ rolls toward its minimum, $\psi_{{}_{\rm min2}}$, the inflaton has an interaction with the extra field, $\chi$, which acts as the preheat field in our setup. Therefore the potential in this setup at the bottom of the true vacuum takes the form
	\begin{equation}
		\label{V-phi-2pt}
		V(\phi)=\frac{1}{2}\left(m^{2}+\lambda'\psi_{{}_{\rm min2}}^{2}\right)\phi^{2}+\frac{1}{2}g^2 \phi^2 \chi^2 \,,
	\end{equation}
	where $m= 1.1162\times10^{-29} m_{{}_{\rm Pl}}$ and $\sqrt{\lambda' \psi_{{}_{\rm min2}}^{2}}= 6.4767\times10^{-30} m_{{}_{\rm Pl}}$. The effective mass resulted will be $\mu\equiv\sqrt{m^{2}+\lambda'\psi_{{}_{\rm min2}}^{2}}= 1.2905\times10^{-29} m_{{}_{\rm Pl}}$ \footnote{$m_{{}_{\rm Pl}}=\sqrt{8\pi} M_{{}_{\rm Pl}}\simeq1.22\times10^{19}\mathrm{GeV}$, where $M_{{}_{\rm Pl}}$ is the reduced Planck mass.}. Also, we assume that the inflaton field remains almost the same as its value at the onset of tachyonic rolling, $\phi_{{}_e}\simeq \phi_{0}= 1.7041\times10^{-13} m_{{}_{\rm P}}$. The model now looks very similar to the model investigated in \cite{Easther:2006vd}, modulo the fact that in our case the amplitude of the inflaton at the start of preheating is much smaller than their case. As we will later elaborate, the model discussed in \cite{Easther:2006vd} cannot satisfy the latest Planck constraints on the CMB scales. We take the preheating coupling, $g$, such that the resonance parameter, $q\equiv g^{2}m_{{}_{\rm Pl}}^2/\mu^{2}=2\times10^{6}$, which is in the center of the instability band.
	
	\begin{figure}
		\begin{center}
			\scalebox{0.65}[0.65]{\includegraphics{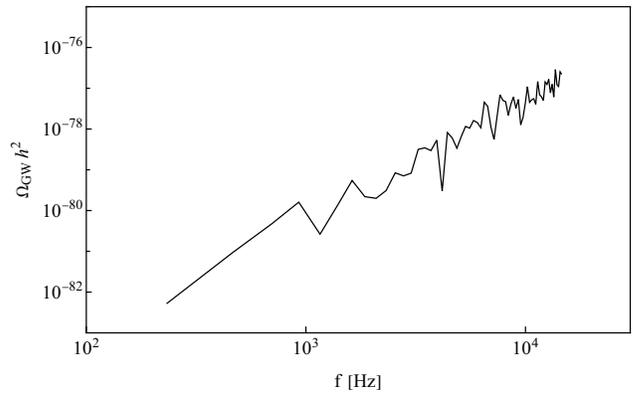}}
			\caption{The results of the LATTICEEASY code for the spectrum of generated SGWB generated from preheating in our scenario with the effective potential \eqref{V-phi-2pt}. In this plot, we have set $\mu = 1.2905\times10^{-29} m_{{}_{\rm Pl}}$, $\phi_{0}= 1.7041\times10^{-13} m_{{}_{\rm Pl}}$, and $q = 2 \times 10^6$. The other relevant parameters in the rescaled units of the code are $N = 128$, $L = 20$, and $t_{{}_{\rm GW}}=3\times10^{4}$.}
			\label{figure:OmegaGW-NANOGravPBHs}
		\end{center}
	\end{figure}
	
	With such values for parameters, we use LATTICEEASY code \cite{Felder:2000hq} to compute the amplitude of  SGWB from preheating in our scenario. The amplitude of the GWs spectrum can be calculated as
	\begin{equation}
		\label{OmegaGWh2}
		\Omega_{{}_{\rm GW}}h^{2}=\Omega_{{}_{\rm r}}h^{2}\frac{d\Omega_{{}_{\rm GW}}(a_{{}_{\rm GW}})}{d\ln k}\left(\frac{g_{{}_0}}{g_{{}_*}}\right)^{1/3}\,,
	\end{equation}
	where $a_{{}_{\rm GW}}$ is evaluated at the end of simulation and $\Omega_{{}_{\rm r}}h^{2}\approx4.3\times10^{-5}$ is the abundance of radiation today. Furthermore, $g_{{}_0}/g_{{}_*}$ is the  the ratio of number degrees of freedom today to the number degrees of freedom at matter-radiation equality, and in our analysis, we take $g_{0}/g_{*}=1/100$. The  quantity $d \Omega_{{}_{\rm GW}} /d\ln k$ is given by
	\begin{equation}
		\label{dOmegaGWdlnk}
		\frac{d\Omega_{{}_{\rm GW}}}{d\ln k}=\frac{1}{\rho_{{}_{\rm crit}}}\frac{d\rho_{{}_{\rm GW}}}{d\ln k}\,.
	\end{equation}
	The result of our simulation is presented in Fig. \ref{figure:OmegaGW-NANOGravPBHs}. In our simulation, we have set the lattice resolution and lattice size as $N = 128$ and $L = 20$, respectively. The spectrum is evaluated at the time $t_{{}_{\rm GW}}=3\times10^{4}$ in the units of the code. We see in the figure that the amplitude of the GWs spectrum in our model is of order $\Omega_{{}_{\rm GW}}h^{2}\sim10^{-82}-10^{-77}$, and its frequency lies in the interval $f\sim10^{2}-10^{4}\mathrm{Hz}$. The frequency range and the resulting amplitude is in a stark difference from the reported NANOGrav signal \cite{NANOGrav:2020bcs}. This proves that the GWs produced from a second-order phase transition at the final stages of inflation in our setup cannot explain the NANOGrav signal.
	
	The result of Ref. \cite{Easther:2006vd}, on the other hand, was suggestive that regardless of the scale of inflation, a fraction proportional to $\Omega_{{}_{\rm GW}}h^{2}\sim10^{-11}-10^{-10}$ of the critical energy density would transform to gravitational waves from preheating, regardless of the scale of inflation. That would be still a bit below the amplitude of the acclaimed NANOGrav signal, but it would be much closer than what we have found.   Let us first briefly review the scenario in \cite{Easther:2006vd}. The authors have analyzed the preheating process at different energy scales using the following potential
	\begin{equation}
		\label{V-phi-chi}
		V(\phi,\chi)=\frac{1}{2}\mu^{2}\phi^{2}+\frac{1}{2}g^{2}\phi^{2}\chi^{2}\,,
	\end{equation}
	where $\phi$ is the inflaton field, $\chi$ is the preheat field, and $g$ denotes the coupling between these fields. The parameter $\mu$ denotes the effective mass of $\phi$  and in the standard quadratic chaotic inflation, it is fixed to be as $\mu\sim10^{-6}m_{{}_{\rm Pl}}$ from the CMB constraints on the amplitude of the scalar power spectrum. However, in the analysis by \cite{Easther:2006vd}, this parameter has been taken as a free parameter. To motivate this assumption, they assume that the full potential during inflation could be written as
	\begin{equation}
		\label{V-phi-chi-sigma}
		 V=\frac{\left(M^{2}-\lambda\sigma^{2}\right)^{2}}{4\lambda}+\frac{m^{2}}{2}\phi^{2}+\frac{h^{2}}{2}\phi^{2}\sigma^{2}+\frac{g^{2}}{2}\phi^{2}\chi^{2}\,.
	\end{equation}
	During the inflationary phase, when $\phi>M/h $, $\sigma$  lies in its false vacuum $\sigma = 0$. During this time, the inflationary potential resulted from \eqref{V-phi-chi-sigma} is
	\begin{equation}
		\label{V-phi-sigma=0}
		V(\phi)=\frac{M^{4}}{4\lambda}+\frac{m^{2}}{2}\phi^{2}\,.
	\end{equation}
	When $\phi=M/h$, $\sigma$ becomes tachyonic and evolves toward its true vacuum at $M/\sqrt{\lambda}$. By assuming $\sigma=\left\langle \sigma\right\rangle$, the potential becomes
	\begin{equation}
		\label{V-phi-M}
		V(\phi)=\frac{1}{2}\left(m^{2}+\frac{h^{2}M^{2}}{\lambda}\right)\phi^{2}\,.
	\end{equation}
	The authors of \cite{Easther:2006vd} argue that one can go to the regime
	\begin{equation}
		\label{m-inequalities}
		m^{2}\ll \frac{h^{2}M^{2}}{\lambda}\,,
	\end{equation}
	in which potential \eqref{V-phi-M} reduces to
	\begin{equation}
		\label{V-phi-mu}
		V(\phi)=\frac{1}{2}\mu^{2}\phi^{2},
	\end{equation}
	\begin{figure*}
		\begin{center}
			\scalebox{1}[1]{\includegraphics{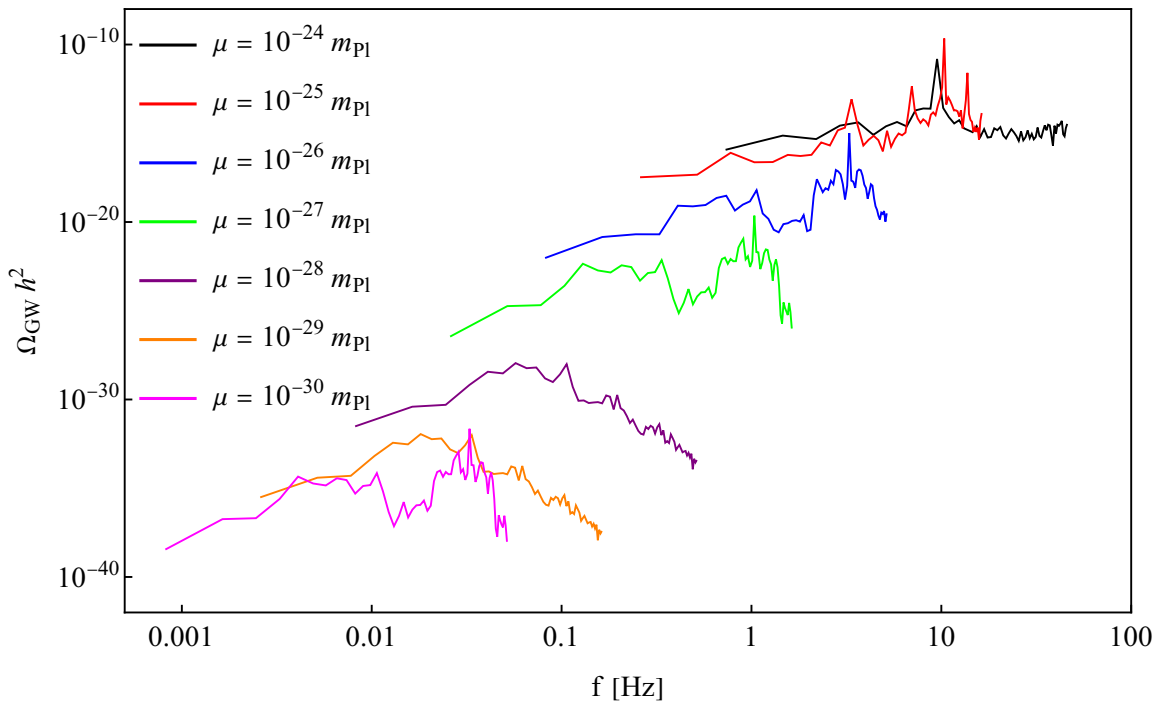}}
			\caption{The results of the LATTICEEASY code for the spectrum of generated GWs from preheating for the effective potential \eqref{V-phi-mu} with different values of $\mu$. Here, we have set $\phi_0 = 0.2\,m_{{}_{\rm Pl}}$ and $q = 2 \times 10^6$. The other relevant parameters in the rescaled units of the code are $N = 128$, $L = 40$, and $t_{{}_{\rm GW}}=1000$.}
			\label{figure:OmegaGW-different_masses}
		\end{center}
	\end{figure*}
	where $\mu^{2}\equiv h^{2}M^{2}/\lambda$ and so can be taken to be a free parameter. However when the inequality \eqref{m-inequalities} is satisfied, and during inflation where $\phi>M/h$, the potential  \eqref{V-phi-M} is dominant by the vacuum energy term. In such a limit, the scalar spectral index is blue, i.e. $n_{{}_{\mathcal{S}}}>1$, which is ruled out by the Planck data \footnote{The model was even ruled out by WMAP three year results \cite{WMAP:2006bqn} which preceded \cite{Easther:2006vd}.}. Nonetheless, in \cite{Easther:2006vd}, the authors have considered $\mu$ as a free parameter and examined values of $\mu$ from $10^{-18}m_{{}_{\rm Pl}}$ to $10^{-6}m_{{}_{\rm Pl}}$. They considered the value of the inflaton field at the start of lattice simulation to be $\phi_0 = 0.2\,m_{{}_{\rm Pl}}$. In their work, they also fixed the resonance parameter as $q\equiv g^{2}m_{{}_{\rm Pl}}^{2}/\mu^{2}=2\times10^{6}$, something that we also assumed in our model above to compute the GW signature from preheating. With these considerations, they computed the stochastic GW  background generated from preheating in their model. They concluded that the amplitude of $\Omega_{{}_{\rm GW}}h^{2}$ spectrum remains always of order $10^{-11}-10^{-10}$ for all the values that they regarded for $\mu$, although the frequency of the spectra for different masses would depend on $\mu$: the smaller the parameter $\mu$, the smaller the peak frequency of the SGWB \footnote{As we will elaborate, Ref. \cite{Easther:2006vd} had claimed to be able to disentangle the energy scale of inflation from preheating using a hybrid model, where the vacuum energy drives inflation but reheating occurs with a massive potential where the mass has nothing to do with the mass of the inflaton during inflation. Still, they have not explained what happens to the energy of the inflaton in the vacuum energy part. In principle, this energy is transformed to the kinetic energy of the field at the beginning of the preheating and potential $\frac{1}{2}\mu^2 \phi_0^2$ is not the total in the field when preheating starts. We continue exploring the effect of lowering the energy scale at the onset of preheating within the setup Ref. \cite{Easther:2006vd} explored. But we do not claim that one can disentangle the energy scale of inflation from that of preheating.}.
	
	The model proposed in \cite{Easther:2006vd} is incapable of reproducing the Planck results at the CMB scales. We will still push the parameter $\mu$ in their model to smaller values to see if the conclusion that the amplitude of stochastic gravitational waves from preheating remains almost constant survives at lower energy scales. Our numerical results for different values of $\mu$ are displayed in Fig. \ref{figure:OmegaGW-different_masses}. To generate this graph, following \cite{Easther:2006vd}, we have considered the value of inflaton field at the start of the simulation as $\phi_0 = 0.2\,m_{{}_{\rm Pl}}$, and the resonance parameter as $q = 2 \times 10^6$. The number of points along each edge of the cubical lattice and the size of the box (i.e. length of each edge) in the lattice units are set here as $N = 128$ and $L = 40$, respectively. The spectrum is evaluated at the simulation time $t_{{}_{\rm GW}}=1000$ in the rescaled units of the LATTICEEASY code \cite{Felder:2000hq}. From the plots presented in Fig. \ref{figure:OmegaGW-different_masses}, we see that for $\mu\gtrsim10^{-25}\,m_{{}_{\rm Pl}}$, the amplitude of GWs spectrum is obtained to be of order $\Omega_{{}_{\rm GW}}h^{2}\sim10^{-11}-10^{-10}$, and thus the results of \cite{Easther:2006vd} remains valid until such energy scales. However, for $\mu\lesssim10^{-26}\,m_{{}_{\rm Pl}}$, the amplitude of spectrum of $\Omega_{{}_{\rm GW}}h^{2}$ saturates at smaller values, indicating a breakdown in the results of \cite{Easther:2006vd} at the relevant energy scales of these masses. Also, from the figure, the spectrum for the lower masses appears at lower frequencies.
	
	The suppression of the amplitude of the SGWB with diminishing the energy scale of preheating might be due to the fact that backreaction effects at lower energy scales kick in at earlier times, well before when the amplitude of SGWB saturates. The study of possible consequences of back-reactions on the dynamics of our two-field setup is out of the scope of the present paper and we leave it for future investigations.

	\section{Primordial Black Holes from Nucleating Bubbles}\label{PBHs}
	
	In this section, we scrutinize the possibility of PBH formation from the collapse of nucleating vacuum bubbles during inflation, during the time interval inflation has ended through the slow-roll violation, and before the field configuration settles in the true vacuum through a first-order phase transition.
	
	As is known in the context of quantum field theory, when the given potential of a theory has two local minima, the false vacuum would be rendered unstable through a barrier penetration due to the quantum fluctuations. During the evolution of such systems, the quantum fluctuations gradually lead to bubbles of true vacuum to materialize and percolate in the sea of the false vacuum \cite{Coleman:1977py}. Bubble expansion and the increase in the number density of bubbles finally cause the system to experience a complete 1PT and settle down in the true vacuum. At the semiclassical level of analysis, the nucleation rate $\gamma$ of true vacuum bubbles is given by \cite{Coleman:1977py,Callan:1977pt}
	\be
	\Gamma= {\cal A} \exp(-S_{{}_{\rm E}})\,,
	\ee
	where $S_{\rm{E}}$ is the Euclidean action calculated at the solution of the corresponding field equation with appropriate boundary conditions and the pre-exponential factor ${\cal A}$ turns out to be of the order  $M^4_{\psi}$ if $M^2_{\psi} > 2H^2$ \cite{Linde:1981zj}. In the case of the quartic potential $V_2$ in (\ref{pot-tot}), one can determine $S_{{}_{\rm{E}}}$ in terms of the parameters of $V_2$ as \cite{Adams:1993zs}
	\begin{equation}
		S_{{}_{\rm E}}= \frac{4 \pi^2}{3 \lambda}(2-\delta)^{-3}(\alpha_1
		\delta+\alpha_2 \delta^2+\alpha_3 \delta^3) \,,
		\label{Euclidian-act}
	\end{equation}
	where $\alpha_1= 13.832 ,~\alpha_2=-10.819,~\alpha_3=2.0765$, and $\delta$ is given by
	\begin{equation}\label{delta}
		\delta=\frac{9 \lambda \alpha}{\gamma^2}+\frac{9 \lambda \lambda'
			\phi^2}{\gamma^2 M^2} \,,
	\end{equation}
	which is valid for $0 < \delta < 2$ or more precisely when the rolling field is larger than its critical value, i.e. $\phi_{{}_{\rm cr}}$. One can also define the probability of false to true vacuum transition as $p=\Gamma/H^4$ and use that to determine the duration in which the 1PT would be accomplished. It has been understood that if this probability exceeds $p_{\rm{c}}\approx 0.24$, because of the effective percolation of the true vacuum bubbles, the universe experiences a complete 1PT \cite{Guth:1982pn}.
	
	In the model at hand, during inflation because of the existence of another direction with smaller vacuum energy, those true vacuum bubbles with radii often smaller than $\mathcal{O}(H^{-1})$ form and then stretch to the much larger lengths due to the exponential expansion of the universe. Immediately after inflation stops by the violation of slow-roll condition, due to bubbles losing their energy on a very short time scale, $\Delta t \sim \frac{H_t}{M_{\rm Pl}^2}\sim 10^{-72} H_t^{-1}$, where $H_t$ is the Hubble parameter in the true vacuum, the {\it{subcritical}} bubbles with radius roughly smaller than $H_t^{-1}$ will collapse to black holes of masses $ 4\pi H_f^{-1}\lesssim M_{\rm{PBH}} \lesssim 32\pi H_f^{-1}$ \cite{Deng:2016vzb,Ashoorioon:2020hln}. On the other hand, the {\it supercritical} bubbles with radii larger than $\mathcal{O}(H_t^{-1})$ keep inflating even after inflation ends due to their large inner vacuum energy. The resulting inflating baby universe would be connected to the parent radiation dominated FRW universe after inflation via a wormhole \cite{Deng:2016vzb} and finally pinches off on the time scales about $t_{{}_{\rm pinch}}\approx M_{{}_{\rm{PBH}}}/M^2_{{}_{\rm{Pl}}}\sim H_t^{-1}(t_{{}_p})$ and then two black holes at the two mouths of wormhole form. Since the time scale of first-order phase transition is $\beta^{-1}\simeq 0.045 H_t^{-1}$, the above inflationary model does not allow for the formation of PBHs from supercritical bubbles although the subcritical ones collapse almost instantaneously.
	
	As computed in \cite{Deng:2016vzb,Ashoorioon:2020hln, Khlopov:2000js}, the mass fraction of the subcritical PBHs is related to the probability of nucleation from the false vacuum to the true one during inflation,
	\begin{equation}\label{approx-mass-fun}
		f(M)\sim B p(t_n){\cal M}_{{}_{\rm eq}}^{1/2} M_{\ast}^{-1/2} \,, M_{{}_{\rm min}}<M<M_{{}_{\ast}}\,,
	\end{equation}
	where $B\sim 10$, and $M_{\ast}\sim 32\pi M_{\rm P}^2 H_t^{-1}$, and ${\cal M}_{{}_{\rm eq}}\sim 10^{17}M_{\odot}$ is of the order of the cold dark matter mass in the Hubble radius at the equality time. The lower bound, $M_{{}_{\rm min}}= 4\pi M_{{}_{\rm Pl}}^2 H_t^{-1}$ comes from the shape fluctuations of the subcritical bubbles that can become large enough for small bubbles preventing them from collapsing to PBHs. As mentioned in section (\ref{sec:results}), regarding the NANOGrav signals, the Hubble scale of 1PT can only take values within  $6.00 \times 10^{-42}\lesssim H_{{}_{\rm pt}} \lesssim 9.22\times 10^{-40} M_{\rm Pl}$. This implies that the mass for those of PBHs which can be generated through such mechanism ranges from $\sim 30$ to $\sim 36000$ solar mass. This not only covers the LIGO mass PBHs, but also the heavier ones. For the first model which exits inflation with 1PT and its peak frequency would fall into the NANOGrav region, the subcritical mass range is $337~M_{\odot}\lesssim M_{{}_{\rm PBH}}\lesssim 2700~M_{\odot}$. For the other specific example that we worked out in section \ref{sec:results} and only the falling tail is in the NANOGrav region, the mass range of produced PBHs is  $2930 \lesssim M_{\rm{PBH}} \lesssim 23441$ solar mass corresponding to $H_{\rm pt}\approx 9.2 \times 10^{-42}M_{\rm{Pl}}$.
	
	One should notice that from the empirical bounds that exist on the mass fraction of PBHs in this mass range, one can constrain the probability of bubble nucleation during inflation. The most stringent constraints are coming from the effect of accreting PBHs on the CMB \cite{Sasaki:2018dmp, Ali-Haimoud:2016mbv, Poulin:2017bwe, Gaggero:2016dpq}. The constraints are highly dependent on the mass of the primordial black holes, but in the mass range that we are focused on in this model, ${\rm few}\times 10~M_{\odot}\lesssim M_{{}_{\rm PBH}}\lesssim{\rm few}\times 10^4~M_{\odot}$, the most restrictive upper bound on the abundance is $f(M_{{}_{\rm PBH}})\lesssim 10^{-6}$. The heavier the mass of the PBHs, the smaller the upper bound on their abundance is. This upper bound on the abundance of the PBHs will translate to an upper bound on the probability of transition from the false ``valley'' to the true one. For the first model with the peak frequency of the gravitational wave in the NANOGrav region and mass of PBHs in the range $337~M_{\odot}\lesssim M_{{}_{\rm PBH}}\lesssim 2700~M_{\odot}$, the nucleation rate of true vacuum bubbles is such that the abundance of the PBHs turns out to be $f(M) \sim 6\times 10^{-13}$ in the corresponding mass range, which is much smaller than the upper bound set from accretion to PBHs. We tried different sets of parameters for the barrier of the potential, $V_2(\phi,\psi)$, but in all those cases, demanding that the SGWB signal falls in the NANOGrav sensitivity band, would suppress the abundance of the subcritical PBHs formed in the mass range ${\rm few}\times10~M_{\odot}$-${\rm few}\times10000~M_{\odot}$ to the level $\lesssim 10^{-10}$. We think this is the characteristic of the quartic potential $V_2(\phi,\psi)$. If we could compute the Euclidean action for a barrier with a more arbitrary function, we believe that it would have been possible to obtain higher nucleation rates such that a larger abundance of the PBHs, along with an SGWB in the NANOGrav region, could be obtained. Noting that the initial abundance of the PBHs is dependent on the nucleation rate from the false vacuum to the true one, one might think that by determining the initial mass function of the PBHs, one can in principle chart the landscape around us. However the mass function of PBHs is known to evolve with time due to mergers, accretion, and possibly spatial clustering \cite{Nakamura:1997sm, Ioka:1998nz, Raidal:2018bbj}. This would make charting the landscape difficult, if not impossible. Nonetheless by modeling these processes one can in principle put bounds on the initial abundance of PBHs during formation and hence the nucleation rate from the false vacuum to the true one.
	
	One might think that the universe is inhomogeneous at the beginning of nucleosynthesis, noting that the time lapse from the end of inflation and the beginning of nucleosynthesis is smaller than when the scale of inflation is high. However, we should note that the phase transition completes in the time scale of $\beta^{-1}$ which is a small fraction of the Hubble time, $H^{-1}$. During the phase transition the universe at most evolves like a matter or radiation-dominated universe. Hence the scale factor has evolved like $a\propto\left((t_{e}+t_{\mathrm{pt}})/t_{e}\right)^{w}$, where $w=1/2$ or $2/3$ (or even $w\simeq 1$). Here, $t_e$ is the time of the end of inflation which is at least $N_e H^{-1}$. Hence, the scale factor would have only evolved slightly, during which the temperature would get uniform across the whole Hubble patch which develops to be the universe today.
	
	If settling to the true vacuum through a second-order phase transition and preheating could lead to the generation of NANOGrav signal since the time scale for preheating could become larger than the time scale during which the supercritical PBHs form, one would expect to not only see the subcritical branch of the PBH mass function but also the supercritical ones, with the mass fraction which would decay like $M^{-1/2}$ for $M_{{}_{\rm PBH}}\geq M_{\ast}$. However, as we noticed above, with the reduction of the inflationary scale required to cover the frequency range in which the NANOGrav probe has claimed to see the SGWB, the amplitude of the GW signal reduces significantly.
	

	
	\section{Concluding Remarks}
	\label{conclusions}
	
	In this paper, we proposed an end of inflation scenario for the SGWB recently reported by the NANOGrav collaboration after the reanalysis of its 12.5-year data \cite{NANOGrav:2020bcs}. The characteristics of the signal seemed to match with the stochastic signal left after a first-order phase transition. Since the energy scale and the reheating temperature from such a phase transition would be above the nucleosynthesis temperature, $T\geq 1$ MeV, we could construct a double field inflationary model \cite{Adams:1993zs}, which satisfies the Planck 2018 constraints at large scales, exit inflation via the violation of slow-roll, but settles to the true vacuum via a first-order phase transition, a process through which the NANOGrav signal is generated. During inflation, another metastable valley develops, and bubbles of true vacuum form during inflation. Those bubbles collapse during the ensuing matter or radiation-dominated universe to form primordial black holes, a correlated signal with the NANOGrav one would be PBHs within the mass range of few tens to few ten thousand of solar mass, depending on the exact energy scale of inflation. The abundance of these PBHs depends on the nucleation rate from the false to true vacuum during inflation. Hence by studying the abundance of such PBHs, we can gain information about the structure of vacua around us \cite{Ashoorioon:2020hln}. Since the process of settling to the true vacuum in this scenario is much smaller than $H_t^{-1}$, the supercritical bubbles whose sizes are bigger than the Hubble radius that have the potential of generating PBHs with mass greater than $M_{\ast}$, will collide before they have the chance to form PBHs. Therefore, we do not see the $M^{-1/2}$ tail, in the mass function for PBHs with a mass larger than the critical mass.

	One may wonder if the merging of the PBHs produced in the mass range of few tens to few ten thousands solar mass, produced from collapsing bubbles, can give rise to a gravitational wave spectrum whose low-frequency tail could potentially contribute to the NANOGrav signal. In our case the initial mass function of the PBHs upon formation is small, $f(M)\lesssim 10^{-10}$, which should make the GWs from such mergers tiny. In fact, with a much higher initial mass function, $0.001\lesssim f(M)\lesssim 0.01$, such a spectrum is known to peak around the advanced LIGO frequency band, $1$ Hz to $10^3$ Hz and the low-frequency tail of such a spectrum falls below the sensitivity of the NANOGrav probe \cite{Raidal:2017mfl} in the relevant frequencies. Hence even if the mergers, accretion, and spatial clustering enhance the initial mass function, still the resulting GW spectrum cannot contribute to the NANOGrav signal.
	
	In the literature, there were claims that the stochastic resonance at the end of inflation, can also generate an SGWB roughly comparable in size and shape with the one generated through a first-order phase transition at the end of inflation \cite{Easther:2006vd}. Hence, we tried to see if a similar SGWB signal could be realized from a second-order phase transition after the end of inflation in our double-field setup. We realized that the amplitude of the GW signal generated from our model, which satisfies the constraints at the CMB  scales is of order, $10^{-82}\lesssim \Omega_{{}_{\rm GW}}h^{2}\lesssim 10^{-77}$, in the frequency band $10^{2}\lesssim f \lesssim 10^{4}$ Hz, which does not match the NANOGrav signal, neither in amplitude nor in frequency. We tried to investigate the issue in the context of the model analyzed in  \cite{Easther:2006vd} too. We noticed the independency of the gravitational wave amplitude from the energy scale of inflation/preheating breaks down at the energy scale $M\lesssim 10^{-14}~m_{{}_{\rm Pl}}$ in that model too, corresponding with the $\mu\lesssim 10^{-26}~m_{{}_{\rm Pl}}$, which is the mass of the field around the true vacuum. A potential reason behind the suppression of the amplitude of the generated SGWB could be that the time scale of the backreaction that shuts off the preheating becomes much smaller than the time needed for the SGWB from preheating to saturate at smaller energy scales. We leave the study of this observation to future studies.

	
	\subsection*{Acknowledgements}
	We thank G. Felder, R. Easther, and J. T. Giblin for helpful discussions. We also thank K. Freese and in particular M. Winkler for detecting a flaw in the numerical prefactor in our formula for the peak frequency, which changed the energy scale in our model. This project has received funding/support from the European Union's Horizon 2020 research and innovation programme under the Marie Sk\l{}odowska-Curie grant agreement No 860881-HIDDeN.

	
	

	
\end{document}